 \documentclass[smallabstract,smallcaptions]{dccpaper}

\usepackage{epsfig}
\usepackage{amsmath}
\usepackage{bm}
\usepackage{amssymb}
\usepackage{color}
\usepackage{makecell}
\usepackage{url}
\usepackage{tikz}
\usepackage{pgfplots}
\usepackage[font=small,skip=1pt]{caption}
\usepackage{subcaption}          
\usepackage{titlesec}
\usepackage[htt]{hyphenat}
\usepackage{pgfplotstable}

\titlespacing*{\paragraph}{0pt}{3.25ex}{0.75ex}
\usepackage{float}
\usepackage{booktabs}
\usepackage{multirow}
\newlength{\figurewidth}
\newlength{\smallfigurewidth}

\newcommand{\img}{\mathbf{x}_t}
\newcommand{\sysout}{\hat{\mathbf{x}}_t}
\newcommand{\latent}{\mathbf{y}}
\newcommand{\qlatent}{\hat{\latent}}
\newcommand{\synthparam}{\bm{\theta}}
\newcommand{\armparam}{\bm{\psi}}
\newcommand{\upparam}{\bm{\upsilon}}
\newcommand{\synth}{f_{\synthparam}}

\newcommand{\upsample}{f_{\upparam}}
\newcommand{\pred}{\bm{\tilde{x}}_t}
\newcommand{\predictionmode}{\bm{\alpha}}
\newcommand{\bidirectionalweight}{\bm{\beta}}
\newcommand{\residue}{\bm{r}}
\newcommand{\refone}{\hat{\mathbf{x}}_{\mathrm{ref}_1}}
\newcommand{\reftwo}{\hat{\mathbf{x}}_{\mathrm{ref}_2}}
\newcommand{\vt}{\bm{v}_{\mathrm{ref}_1 \rightarrow t}}
\newcommand{\vtwo}{\bm{v}_{\mathrm{ref}_2 \rightarrow t}}

\definecolor{darkpastelpurple}{rgb}{0.59, 0.44, 0.84}
\definecolor{myred}{HTML}{E76F51}
\definecolor{myblue}{HTML}{376996}
\definecolor{mygreen}{HTML}{2A9D8F}
\definecolor{mypurple}{HTML}{822E81}
\definecolor{mywhite}{HTML}{f0dfbb}

\begin{document}

\title
{\large
    Cool-chic video:~Learned video coding with 800 parameters
}

\author{%
Thomas Leguay$^{\ast}$, Théo Ladune$^{\ast}$, Pierrick Philippe$^{\ast}$, Olivier Déforges$^{\dag}$\\[0.5em]
{\small\begin{minipage}{\linewidth}\begin{center}
\begin{tabular}{ccc}
$^{\ast}$Orange Innovation & \hspace*{0.5in} & $^{\dag}$IETR \\
Rennes, France && Rennes, France\\
\url{firstname.lastname@orange.com} && \url{olivier.deforges@insa-rennes.fr}
\end{tabular}
\end{center}\end{minipage}}
}

\Support{Submitted to DCC 2024 in the course of the Challenge on Learned Image Compression (CLIC)\\
© 2024 IEEE.  Personal use of this material is permitted.  Permission from IEEE must be obtained for all other uses, in any current or future media, including reprinting/republishing this material for advertising or promotional purposes, creating new collective works, for resale or redistribution to servers or lists, or reuse of any copyrighted component of this work in other works.}

\maketitle
\thispagestyle{empty}
\vspace*{0.5cm}
\begin{abstract}
We propose a lightweight learned video codec with 900 multiplications per
decoded pixel and 800 parameters overall. To the best of our knowledge, this is
one of the neural video codecs with the lowest decoding complexity. It is built
upon the overfitted image codec Cool-chic and supplements it with an inter
coding module to leverage the video's temporal redundancies. The proposed model
is able to compress videos using both low-delay and random access configurations
and achieves rate-distortion close to AVC while outperforming other overfitted
codecs such as FFNeRV. The system is made open-source:
\url{orange-opensource.github.io/Cool-Chic}.
\end{abstract}

\section{Introduction \& related works}

\paragraph{Conventional codecs.} Conventional video codecs such as AVC
\cite{avc}, HEVC \cite{hevc} and VVC \cite{VVC} dominate the video codec
market. These codecs follow a similar paradigm where the encoder has many coding
modes available (\textit{e.g.} for the partitioning, prediction or transform)
and selects the most suited ones for the current image based on a
rate-distortion optimization. At the decoder-side only the selected
tools are applied to recover the decoded image, ensuring low decoding
complexity.


\paragraph{Autoencoders.} Neural video codecs \cite{li23} obtain performance
close to state-of-the-art conventional codecs such as VVC. These systems are
built on autoencoders whose parameters are optimized at training time following
a rate-distortion loss function. When actually compressing a video
(\textit{i.e.} at inference time), these parameters are used \textit{as is} with
no further rate-distortion optimization. Since the same parameters are applied
identically on all signals, they have to be numerous to deal with every possible
signal. As such, the decoding complexity of autoencoder-based codecs can
reach up to a million multiplications per decoded pixel \cite{li23}. Despite
impressive rate-distortion results, this complexity might hinder the real-world
usage of neural codecs.


\paragraph{Overfitted codecs.} Overfitted codecs such as NeRV \cite{nerv}, HNeRV
\cite{hnerv}, FFNeRV \cite{ffnerv} and HiNeRV \cite{hinerv} propose to encode a
video as a neural representation. During the encoding, a neural network mapping
pixel coordinates to pixel values is trained to minimize the rate-distortion
tradeoff of the video. Decoding involves inferring the trained neural
representation. Overfitted codecs bridge the gap between conventional and
autoencoder-based codecs by reintroducing video-wise rate-distortion
optimization. This allows overfitted codecs to maintain a reasonable decoding
complexity while still retaining the benefits of learned codecs. However, we
believe that current overfitted approaches have two drawbacks.
\begin{enumerate}
    \item Video compression is framed as a neural network compression. We argue
    this makes the redundancy exploitation harder. Indeed, videos tend to have
    easily exploitable spatiotemporal redundancies whereas neural networks are not
    as clearly organized, making their compression more cumbersome.
    \item To amortize the relatively important rate of the network, a single
    neural representation is used for the entire video. This forbids the usage
    of low-delay coding configurations since several hundreds of frames are
    encoded together.
\end{enumerate}

\paragraph{Cool-chic.} Unlike other overfitted codecs, the Cool-chic image codec
\cite{coolchic, mmsp} relies on overfitted latent feature maps as the compressed
representation of an image. While overfitted neural networks are still used
(\textit{e.g.} to map the latent variables back to the image domain), their size
and complexity are significantly reduced. Moreover, the latent feature maps are
spatially organized, easing the exploitation of spatial redundancies. As a
result, Cool-chic offers image compression performance on par with HEVC while
requiring less than 1~000 multiplications per decoded pixel \cite{mmsp}.


\paragraph{Contributions.} This paper builds upon Cool-chic to design a
low-complexity learned video codec, addressing the drawbacks of previous
overfitted codecs. In particular, temporal redundancies are exploited through
motion compensation. This allows successive video frame to be encoded
individually (\textit{i.e.} low-delay settings) instead of compressing hundreds
of frames together to amortize the neural network rate. This is achieved by
adding an inter coding module after a Cool-chic decoder to perform motion
compensation and residual coding. Our contributions are as follows:


\begin{itemize}
    \item Low-complexity decoder with 900 multiplications per decoded pixel;
    \item Frame-wise encoding, enabling both low-delay and random access encoding;
    \item Rate-distortion results on par with AVC and better than previous overfitted codec (FFNeRV);
    \item The model is made open-source to foster further low-complexity neural compression investigation.
\end{itemize}


\section{Method}

\begin{figure}[t]
    \includegraphics[width=6in]{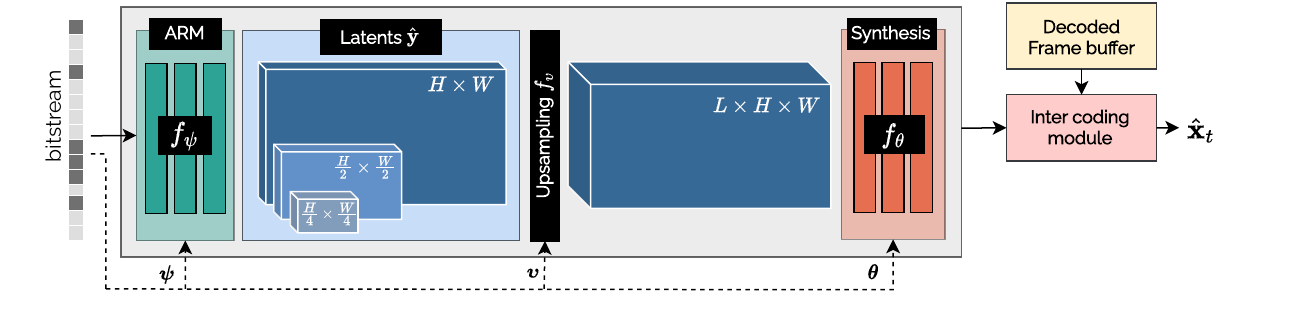}
    \caption{Decoding of a video frame.}
    \centering
    \label{fig:cool-chic-decoding}
\end{figure}

This paper proposes to extend the open-source image Cool-chic \cite{coolchic} to
an inter-frame coding task. In particular, it considers the compression of an $H
\times W$ video frame $\img$ with up to two reference frames $\left(\refone,
\reftwo\right)$ available at the decoder. To this end, Cool-chic is supplemented
with an inter coding module, leveraging temporal redundancies of the video.

\subsection{Background: Cool-chic image codec}

This section summarizes the Cool-chic image coding scheme on top of which the
proposed method is built. More details can be found in \cite{coolchic}.

\paragraph{Decoding.} Figure \ref{fig:cool-chic-decoding} presents the decoding
process of Cool-chic. First, neural network parameters are retrieved from the
bitstream using an entropy coding scheme. Then, an auto-regressive probability
model parameterized by the MLP $f_{\armparam}$ drives the entropy decoding of a
hierarchical set of $L$ two-dimensional latent variables $\qlatent$. The latent
variables are subsequently upsampled through the neural network $\upsample$
yielding a dense $L \times H \times W$ latent representation. Finally, the
synthesis MLP $\synth$ maps the $L \times H \times W$ latents to the decoded
features \textit{e.g.} the YUV channels $\hat{\mathbf{x}}$.

\paragraph{Encoding.} The encoding objective is to optimize the neural
network parameters as well as the latent variables to minimize the
rate-distortion cost of the video frame:
\begin{equation}
    \qlatent, \armparam, \upparam, \synthparam = \arg\min \mathrm{D}(\mathbf{x},\hat{\mathbf{x}}) + \lambda \mathrm{R}(\qlatent),
    \label{eq:rd_cost_encoding_image}
\end{equation}
where $\mathrm{D}$ denotes a distortion metric \textit{e.g.} the MSE, and
$\mathrm{R}$ the rate. At the end of the encoding stage, the parameters
are entropy coded into the bitstream.

\subsection{Contribution: inter coding module}

\begin{figure}[t]
    \centering
    \includegraphics[width=5in]{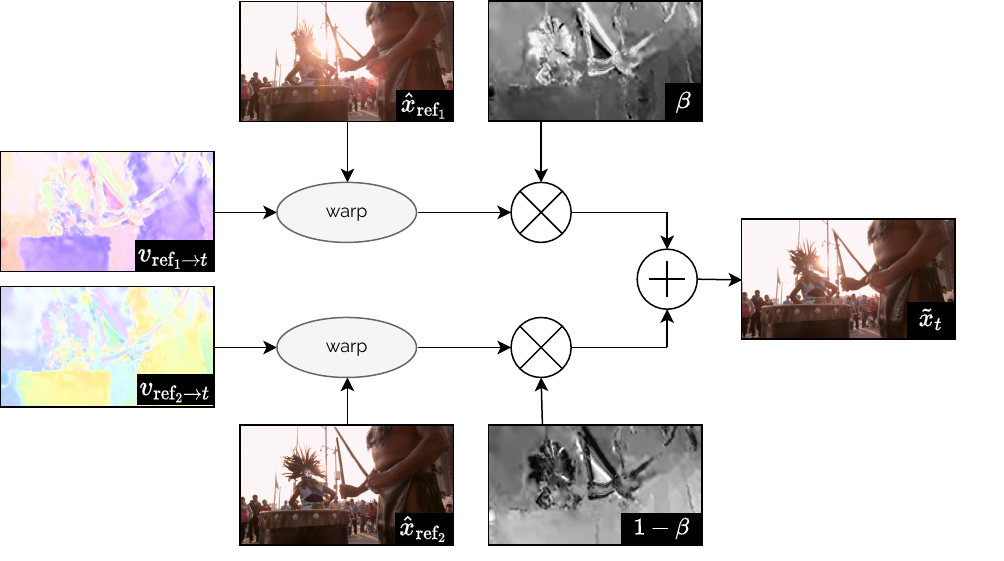}
    \caption{Bidirectional weighted motion compensation. In the prediction
    weighting $\bidirectionalweight$, black corresponds to 0 and white to 1.}
    
    \label{fig:motion-compensation}
\end{figure}

Inter frame coding is achieved by adding an inter coding module after the
decoding process previously described. This module implements both a motion
compensation step and a residual coding to leverage temporal redundancies. All
the input variables required by the inter coding module are obtained by changing
the number of output features of the synthesis $\synth$. Consequently, the
synthesis output for a B-frame is
\begin{equation}
    \synth(\upsample(\qlatent)) = \{\vt, \vtwo, \residue, \predictionmode, \bidirectionalweight\}.
\end{equation}
The following details the roles of the different quantities computed by the synthesis.

\paragraph{Motion compensation.} The inter coding module performs a
bidirectional weighted motion compensation (Fig. \ref{fig:motion-compensation}).
Both references $\left(\refone, \reftwo\right)$ are warped using two dense
optical flows $\left(\vt, \vtwo\right)$. Then, a pixel-wise continuous weighting
$\bidirectionalweight \in \left[0, 1\right]^{H \times W}$ is applied to blend
the two warpings, yielding the temporal prediction $\pred$:
\begin{equation}
    \pred = \bidirectionalweight \odot \mathrm{warp}(\refone,\vt) + (1-\bidirectionalweight) \odot \mathrm{warp}(\reftwo,\vtwo),
\end{equation}
where $\odot$ denotes element-wise multiplication.

\paragraph{Residual coding \& prediction mode.} Since a prediction $\pred$ is
available at the decoder, the system conveys a residue $\residue$ to
correct the prediction. Yet, cases might arise where the prediction becomes
counter-productive (\textit{e.g.} disocclusions). To handle such situations, a
continuous pixel-wise prediction mode selection $\predictionmode \in \left[0,
1\right]^{H \times W}$ is introduced. Its role is to mask the prediction when
required. The decoded frame $\sysout$ is obtained by combining the masked
prediction with the residue:
\begin{equation}
    \sysout = \residue + \predictionmode \odot \pred.
\end{equation}

\begin{figure}[t]
    \includegraphics[width=6in]{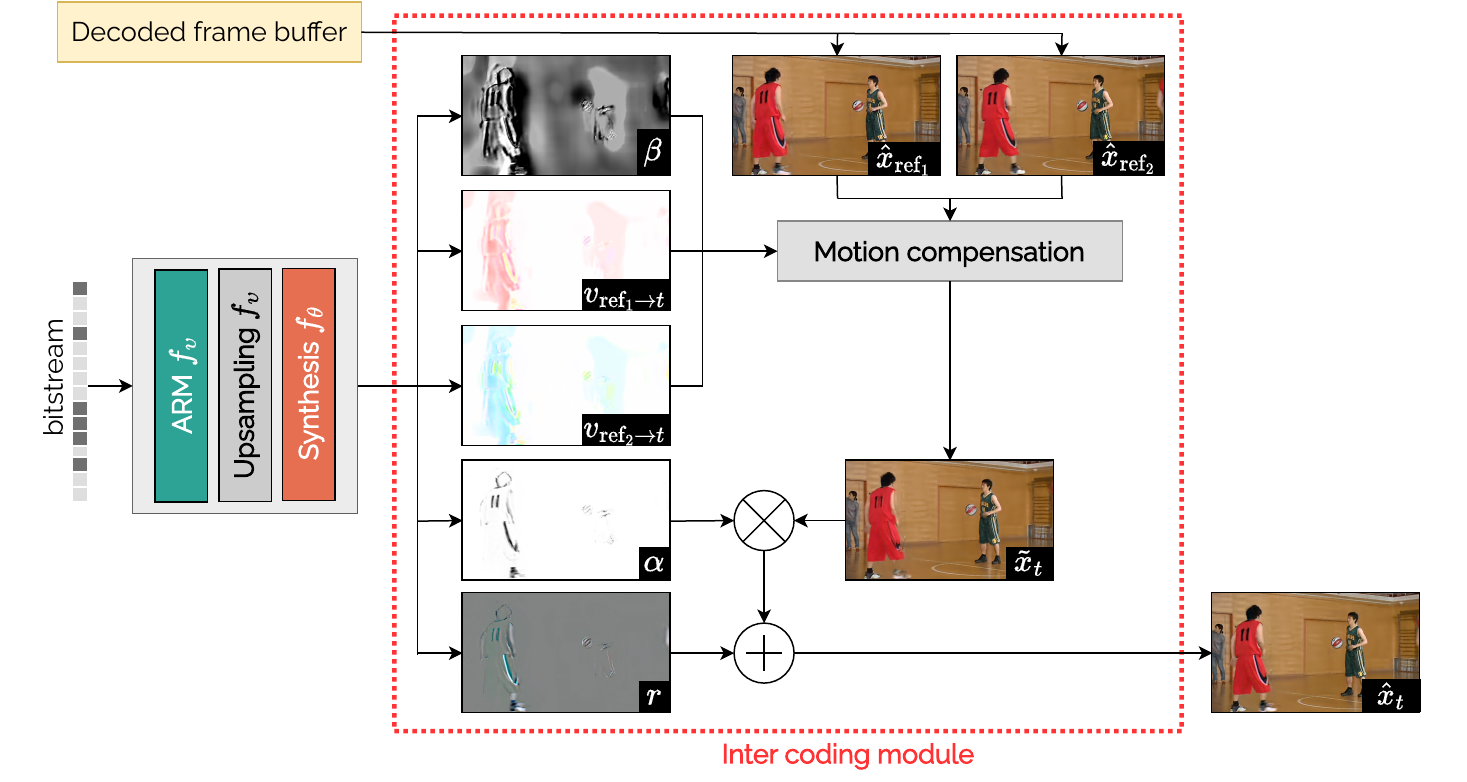}
    \caption{Decoding a B-frame using the inter coding module}
    \centering
    \label{fig:decoding-a-video-frame}
\end{figure}

\paragraph{Overall decoding scheme.} Figure \ref{fig:decoding-a-video-frame}
presents the entire decoding process of a video frame. First, the Cool-chic decoder
retrieves the required quantities from the bitstream. Then, the proposed inter
coding module computes a prediction through motion compensation. The prediction
is subsequently masked and corrected by a residue, yielding the decoded frame.
If a frame does not have two references, some signals are disabled:
\begin{itemize}
    \item P-frame (1 reference):~the second optical flow $\vtwo$ is set to 0 and
    the prediction weighting $\bidirectionalweight$ is set to 1;
    \item I-frame (0 reference):~relies solely on the residue to retrieve the
    decoded frame.
\end{itemize}

\paragraph{Encoding.} The encoding of a frame $\img$ looks for the parameters of
the Cool-chic decoder that minimize the rate-distortion cost stated in eq.
\eqref{eq:rd_cost_encoding_image}. Unlike in the original image coding scenario,
the decoded frame is generated by feeding the synthesis output into the inter
coding module. Since the optimization takes this additional module into account,
the Cool-chic decoder learns to output relevant signals \textit{i.e.} optical
flows, prediction weighting, prediction mode and residue.

\section{Experimental results}

\subsection{Rate-distortion results}

\paragraph{Test conditions.} Experiments are carried out on the CLIC24
validation dataset \cite{clic24dataset}, composed of 30 high-resolution videos
with framerates ranging from 24 to 60 fps. In order to reduce the encoding time,
only the first 33 frames of each video are compressed. Two coding configurations
are investigated:~Random Access (RA) composed of one intra followed by a
hierarchical GOP of 32 frames and Low-delay P (LDP) composed of one intra and 32
P-frames. The MSE is computed in the YUV 4:2:0 domain \textit{i.e.}
$\mathrm{MSE}(\sysout,\img) = \frac{4}{6} \mathrm{MSE_Y}(\sysout,\img) +
\frac{1}{6} \mathrm{MSE_U}(\sysout,\img) + \frac{1}{6}
\mathrm{MSE_V}(\sysout,\img)$. The average PSNR is computed by averaging the MSE
weighted by the number of pixels for each sequence. The average rate in Mbits/s
is obtained by averaging the framerate.

\paragraph{Anchors.} Cool-chic video performance is assessed against two
conventional anchors (AVC and HEVC) through the x264 and x265 encoders. The
following command lines are used to encode the videos, with \texttt{H} and
\texttt{W} the video size, \texttt{QP} the quality command and \texttt{codec}
either x264 or x265:
\begin{itemize}
    \item Random Access:~\footnotesize{\texttt{ffmpeg -f rawvideo -pix\_fmt yuv420p -s <W>x<H>
    -i video.yuv \\ -c:v libx<codec> -x<codec>-params
    "keyint=64:min-keyint=64" -crf <QP>\\ -preset medium -tune psnr  -frames:v 33
    out.bin}}
    \item \normalsize Low-delay P:~\footnotesize{\texttt{ffmpeg -f rawvideo -pix\_fmt yuv420p -s <W>x<H>
    -i video.yuv -c:v \\ lib<codec> -<codec>-params
    "bframes=0:keyint=64:min-keyint=64" -crf <QP>\\ -preset medium -tune psnr  -frames:v 33
    out.bin}}
\end{itemize}

\paragraph{Decoder architecture.} The decoder architecture is identical for all
frames, except for the last synthesis layer whose size varies with the number of
signals required by the inter coding module. 7 hierarchical latent feature maps are
used with a resolution ranging from $H \times W$ to $\tfrac{H}{64} \times
\tfrac{W}{64}$. Table \ref{tab:config} summarizes the architecture of the
different Cool-chic modules. Note that the proposed inter coding module has no
learned parameters:~it is only used to combine the different output features of
Cool-chic.

\begin{table}[H]
    \captionsetup{skip=10pt} 
    \centering
    \small
    \begin{tabular}{c||c|c|c|c||c}
                                                    & ARM               & Upsampling            & Synthesis    & Inter   & \multirow{2}{*}{Total}\\
                                                    & $f_{\armparam}$   &  $f_{\upparam}$       & $f_{\synthparam}$ & coding & \\
                                                    \midrule
        \multirow{3}{*}{\small Architecture}        & Lin 12 / 12            &                               & Lin 7\hphantom{$0$}  / 18  &               & \\
                                                    & Lin 12 / 12            &  TConv $k = 8\ s = 2$         & Lin 18 / $X$               &               & \\
                                                    & Lin 12 / 2\hphantom{4} &                               & Lin $X$ / $X$              &               & \\
        \midrule
        \small Parameters                           & 338                       & 64                            & 405                           & 0             & 807\\
        \midrule
        \small Multiplication                       & \multirow{2}{*}{415}      & \multirow{2}{*}{130}          & \multirow{2}{*}{369}          & \multirow{2}{*}{10}            &\multirow{2}{*}{924}     \\
        \small per decoded pixel & & & & \\
        \midrule
        \end{tabular}
    \caption{Decoder architecture. \textit{Lin $I/O$} denotes a linear feature
    with $I$ input and $O$ output features. All layers are followed by a ReLU,
    except the last one of each module. TConv is a transpose convolution of
    stride 2 and kernel size 8. $X$ denotes the number of output features from
    the synthesis:~3 for I-frames, 6 for P-frames and 9 for B-frames. Parameter
    count and complexity are given for the worst case \textit{i.e.} $X=9$.}
    \label{tab:config}
\end{table}

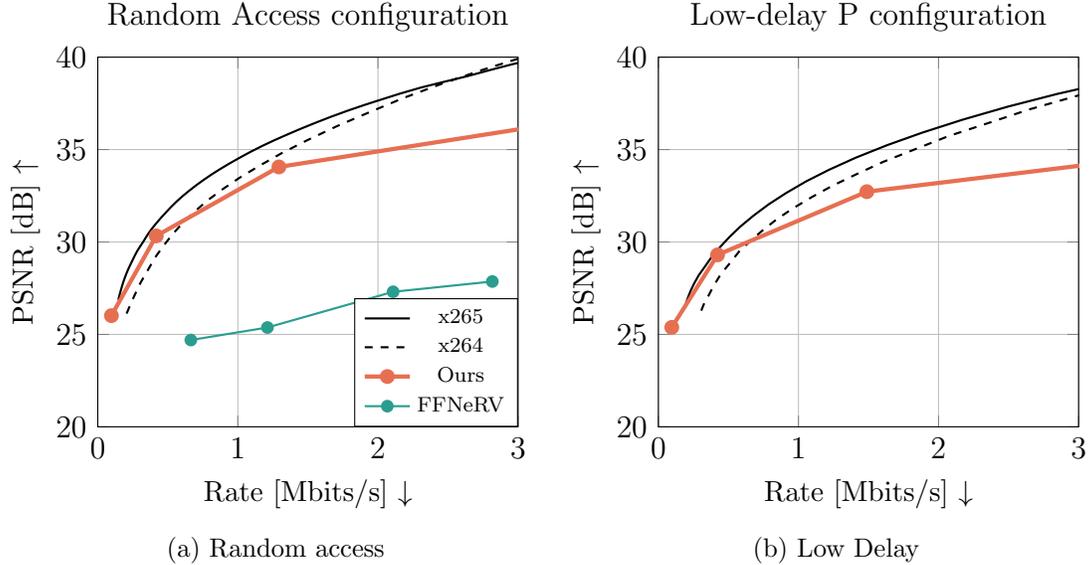
\begin{figure}
    \centering
    \begin{subfigure}{0.48\linewidth}
    \begin{tikzpicture}
        \begin{axis}[
            grid= both,
            xlabel = {\small Rate [Mbits/s] $\downarrow$},
            ylabel = {\small PSNR [dB] $\uparrow$ } ,
            xmin = 0., xmax = 3.,
            ymin = 20, ymax = 40,
            ylabel near ticks,
            xlabel near ticks,
            width=0.98\linewidth,
            height=6.5cm,
            xtick distance={1.},
            ytick distance={5},
            ylabel style={yshift=-2pt},
            minor y tick num=0,
            minor x tick num=0,
            legend style={at={(1.0,0.)}, anchor=south east},
            title={Random Access configuration},
        ]

            \addplot[thick, black, mark=none, smooth] table [x=rate_mbs,y=psnr_db] {data/x265.tsv};
            \addlegendentry{\scriptsize x265}

            \addplot[thick, dashed, black, mark=none, mark options={solid}] table [x=rate_mbs,y=psnr_db] {data/x264.tsv};
            \addlegendentry{\scriptsize x264}

            \addplot[ultra thick,  myred, mark=*, mark options={solid}] table [x=rate_mbs,y=psnr_db] {data/average_results.tsv};
            \addlegendentry{\scriptsize Ours}

            \addplot[mygreen, thick, mark=*] table [x=rate_mbs,y=psnr_db] {data/ffnerv.tsv};
            \addlegendentry{\scriptsize FFNeRV}

        \end{axis}
    \end{tikzpicture}
    \caption{Random access}
    \label{fig:average_rd:ra}
    \end{subfigure}
    \begin{subfigure}{0.48\linewidth}
        \begin{tikzpicture}
            \begin{axis}[
                grid= both,
                xlabel = {\small Rate [Mbits/s] $\downarrow$},
                ylabel = {\small PSNR [dB] $\uparrow$ } ,
                xmin = 0., xmax = 3.,
                ymin = 20, ymax = 40,
                ylabel near ticks,
                xlabel near ticks,
                width=0.98\linewidth,
                height=6.5cm,
                xtick distance={1.},
                ytick distance={5},
                ylabel style={yshift=-2pt},
                minor y tick num=0,
                minor x tick num=0,
                title={Low-delay P configuration},
            ]
                \addplot[thick, dashed, mark=none, mark options={solid}] table [x=rate_mbs,y=psnr_db] {data/x264_ldp.tsv};

                \addplot[thick, black, mark=none] table [x=rate_mbs,y=psnr_db] {data/x265_ldp.tsv};

                \addplot[ultra thick,  myred, mark=*, mark options={solid}] table [x=rate_mbs,y=psnr_db] {data/cool-chic-ldp.tsv};
            \end{axis}
        \end{tikzpicture}
        \caption{Low Delay}
        \label{fig:averade_rd:ldp}
        \end{subfigure}
    \caption{Rate-distortion performances on CLIC 2024 dataset. PSNR is computed in the YUV420 domain.}
\end{figure}

\begin{figure*}[ht]
    \centering
    \begin{subfigure}{0.32\linewidth}
        \begin{tikzpicture}
            \begin{axis}[
                grid= both,
                xlabel = {\small Rate [Mbits/s] $\downarrow$},
                ylabel = {\small PSNR [dB] $\uparrow$} ,
                xmin = 0., xmax = 3.,
                ymin = 30., ymax = 50.,
                ylabel near ticks,
                xlabel near ticks,
                width=\linewidth,
                height=5cm,
                xtick distance={1},
                ytick distance={5},
                minor y tick num=0,
                minor x tick num=0,
            ]

            \addplot[thick, dashed, mark=none] table [x=rate_mbs,y=psnr_db] {data/cda_x264.tsv};

            \addplot[thick, solid, mark=none] table [x=rate_mbs,y=psnr_db] {data/cda_x265.tsv};

            \addplot[ultra thick,  myred, mark=*, mark options={solid}] table [x=rate_mbs,y=psnr_db] {data/cda.tsv};

            \addplot[mygreen, mark=*] table [x=rate_mbs,y=psnr_db] {data/cda_ffnerv.tsv};
                
            \end{axis}
        \end{tikzpicture}
        \caption{Sequence cda...}
        \label{fig:seq_cda}
    \end{subfigure}
    \begin{subfigure}{0.32\linewidth}
        \begin{tikzpicture}
            \begin{axis}[
                grid= both,
                xlabel = {\small Rate [Mbits/s] $\downarrow$},
                ylabel = {\small PSNR [dB] $\uparrow$ } ,
                xmin = 0., xmax = 6.,
                ymin = 20, ymax = 40.,
                ylabel near ticks,
                xlabel near ticks,
                width=\linewidth,
                height=5cm,
                xtick distance={2},
                ytick distance={5},
                minor y tick num=0,
                minor x tick num=0,
            ]

            \addplot[thick, dashed, mark=none] table [x=rate_mbs,y=psnr_db] {data/9a7_x264.tsv};

            \addplot[thick, solid, mark=none] table [x=rate_mbs,y=psnr_db] {data/9a7_x265.tsv};

            \addplot[ultra thick,  myred, mark=*, mark options={solid}] table [x=rate_mbs,y=psnr_db] {data/9a7.tsv};

            \addplot[mygreen, thick, mark=*] table [x=rate_mbs,y=psnr_db] {data/9a7_ffnerv.tsv};
                
            \end{axis}
        \end{tikzpicture}
        \caption{Sequence 9a7...}
        \label{fig:seq_9a7}
    \end{subfigure}
    \begin{subfigure}{0.32\linewidth}
        \begin{tikzpicture}
            \begin{axis}[
                grid= both,
                xlabel = {\small Rate [Mbits/s] $\downarrow$},
                ylabel = {\small PSNR [dB] $\uparrow$} ,
                xmin = 0., xmax = 6.,
                ymin = 20, ymax = 40.,
                ylabel near ticks,
                xlabel near ticks,
                width=\linewidth,
                height=5cm,
                xtick distance={2},
                ytick distance={5},
                minor y tick num=0,
                minor x tick num=0,
                legend style={at={(1.0,0.)}, anchor=south east},
            ]

            \addplot[thick, dashed, mark=none] table [x=rate_mbs,y=psnr_db] {data/60c_x264.tsv};

            \addplot[thick, solid, mark=none] table [x=rate_mbs,y=psnr_db] {data/60c_x265.tsv};

            \addplot[ultra thick, myred, mark=*, mark options={solid}] table [x=rate_mbs,y=psnr_db] {data/60c.tsv};

            \addplot[mygreen, thick, mark=*] table [x=rate_mbs,y=psnr_db] {data/60c_ffnerv.tsv};

            \end{axis}
        \end{tikzpicture}
        \caption{Sequence 60c...}
        \label{fig:seq_60c}
    \end{subfigure}
    \caption{Rate-distortion results on 3 CLIC 2024 videos in Random Access configuration.}
    \label{fig:rd-examples-sequence-wise}
\end{figure*}
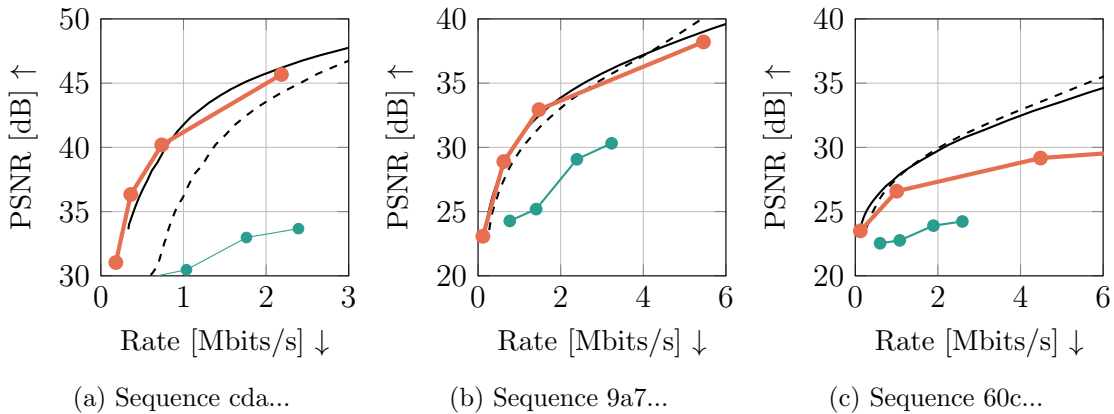


\paragraph{Random Access performance}. Figure \ref{fig:average_rd:ra} presents
the rate-distortion results for the random access configuration. At low rate,
Cool-chic video outperforms x264. However, the proposed
system obtains worse performance at higher rates. For instance, Fig.
\ref{fig:rd-examples-sequence-wise} shows the system results on 3 different
sequences from the CLIC24 dataset. Some sequences (Fig. \ref{fig:seq_cda} and
\ref{fig:seq_9a7}) are better compressed with Cool-chic video than with
conventional codecs. Yet, other sequences (Fig. \ref{fig:seq_60c}) are
challenging for Cool-chic video. We believe that the proposed model struggles to
capture some kind of motion (\textit{e.g.} zoom in sequence 60c) while
successfully estimates the translations present in sequence 9a7.

\paragraph{FFNeRV}. Cool-chic random access performance are also compared to
FFNeRV, an open-source overfitted codec \cite{ffnervCode}. FFNeRV performs worse
than Cool-chic, requiring 10 times more rate to achieve the same quality than
Cool-chic. This is likely due to FFNeRV conveying most of the coded video as
neural network parameters, making it more challenging to exploit spatio-temporal
redundancies.

\paragraph{Low-delay P performance.} Figure \ref{fig:averade_rd:ldp} shows the
performance of the proposed codec for the low-delay P configuration. While low
rate performance is compelling (competitive with x265), higher rates tend to
be worse. Nevertheless, it shows that Cool-chic video is able to perform
low-delay coding thanks to its frame-wise encoding. This is not the case for
other overfitted codecs (\textit{e.g.} FFNeRV) which encode jointly an
important number of consecutive frames.



\paragraph{Complexity. }While Cool-chic video does not yet offer
state-of-the-art compression performance, it stands out for its very low
decoder-side complexity. Table \ref{tab:complexity_comparison} presents the
decoding complexity of different algorithms:~ FFNeRV, an overfitted codec and Li
2023 an  autoencoder-based codec. Both FFNeRV and Cool-chic video present a
significantly lower decoder complexity due to their overfitted nature. Moreover,
the proposed Cool-chic video offers an even lower complexity than FFNeRV. Since
Cool-chic relies on feature maps---not on network weights---to carry information
about the signal, the complexity does not increase with the quality of the
decoded video.

\begin{table}[H]
    \captionsetup{skip=10pt} 
    \centering
    \begin{tabular}{c | c}
        Model & Complexity [kilo multiplication / decoded pixel] $\downarrow$\\
        \midrule
        Cool-chic video & 0.9 \\
        FFNeRV \cite{ffnerv} & From 4.1 (low rate) to 17.8 (high rate) \\
        Li 2023 \cite{li23} & 422\\
    \end{tabular}
    \caption{Decoding complexity comparison of different learned codecs.}
    \label{tab:complexity_comparison}
\end{table}


\subsection{Ablation study}

This section proposes to study the relevance of the different components of the
proposed inter coding module namely the prediction mode $\predictionmode$ and
the bidirectional prediction weighting $\bidirectionalweight$. Results of these
ablations are obtained by measuring the BD-rate (relative rate required for the
same quality) when disabling different components. Table \ref{tab:ablation}
presents the results.

\begin{table}[H]
    \centering
    \small
    \begin{tabular}{c c|c}
        Prediction mode $\predictionmode$  & Bidirectional weighting $\bidirectionalweight$ & BD-rate vs. full model [\%] $\downarrow$\\
        \hline
        \hfill & \checkmark & 85.7 \\
        \checkmark & \hfill & 10.8\\
        \checkmark & \checkmark & 0.0
    \end{tabular}
    \caption{Inter coding module ablation study. Coding configuration is random
    access.}
    \label{tab:ablation}
\end{table}


\paragraph{Prediction mode $\predictionmode$.} The pixel-wise prediction mode
allows some spatial location to ignore the temporal prediction \textit{i.e.} to
resort to plain intra coding when it is beneficial. For content which can not
be easily predicted (\textit{e.g.} disocclusions or scene cuts), intra coding
might offer a better rate-distortion trade-off than conveying an important
prediction error. This is reflected by a rate increase of 85.7~\% when setting
$\predictionmode = 1$ for all pixels.

\paragraph{Bidirectional prediction weighting $\bidirectionalweight$.} The
pixel-wise prediction weighting enables the system to focus on one particular
reference frame for some areas. It is particularly useful when pixels in the
video frame to code are only present in one of the two references. This is
illustrated by a rate increase of 10.8~\% when setting $\bidirectionalweight =
\tfrac{1}{2}$ for the entire frame.


\section{Limitations and future work}

Although Cool-chic video presents interesting coding performance while offering
a low decoding complexity, it suffers from some limitations which need to be
addressed in future work.

\paragraph{Motion estimation.} During the encoding of a video frame
(\textit{i.e.} the training of the decoder), it may be difficult to capture
relevant motion information. This explains the relatively weak performance of
Cool-chic video on videos featuring challenging motions such as zooming effects.
This could be improved through the addition of an auxiliary loss during the
encoding to help learning better motion information.

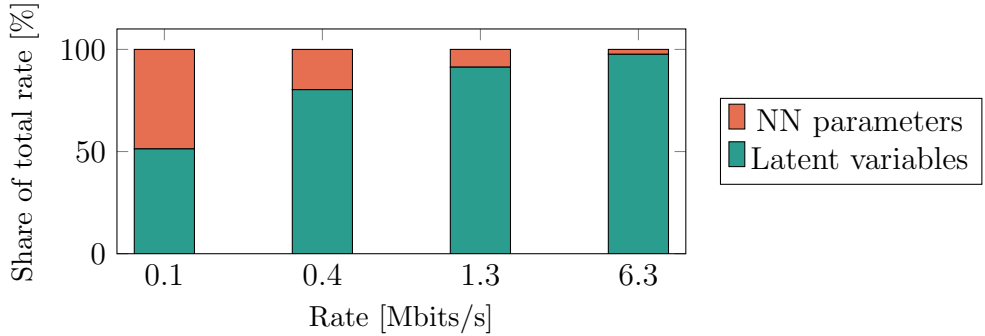
\begin{figure}
    \centering
    \pgfplotstableread{
Label latent_grids_rate nn_parameters_rate
0.1 51.3 48.7
0.4 80.3 19.7
1.3 91.3 8.7
6.3 97.6 2.4
    }\testdata
\begin{tikzpicture}
    \begin{axis}[
        ybar stacked,
        ylabel = {\small Share of total rate [\%]},
        xlabel = {\small Rate [Mbits/s]} ,
        ymin=0,
        ymax=110,
        xtick=data,
        legend style={at={(1.06,0.5)}, anchor=west},
        reverse legend=true,
        xticklabels from table={\testdata}{Label},
        xticklabel style={text width=2cm,align=center},
        height=0.2\textheight,
        width=0.6\textwidth,
        bar width=0.8cm,
    ]
    \addplot [fill=mygreen] table [y=latent_grids_rate, meta=Label, x expr=\coordindex] {\testdata};
    \addlegendentry{Latent variables}
    \addplot [fill=myred] table [y=nn_parameters_rate, meta=Label, x expr=\coordindex] {\testdata};
    \addlegendentry{NN parameters}

    \end{axis}
\end{tikzpicture}
\caption{Relative share of the latent variables and neural network parameters.}
\label{fig:share_rate}
\end{figure}

\paragraph{Neural network rate.} Unlike other overfitted codecs, Cool-chic video
conveys most of the required information through a latent representation instead
of the neural network parameters. Yet, it still has around 800 parameters
which are transmitted for each frame alongside the latent representation. As
shown in Fig. \ref{fig:share_rate} the share of the neural network rate can reach up to 50~\%
for the lower rates. Reducing this rate by leveraging the temporal redundancies
of the networks across a video would help reduce the rate of the networks.

\paragraph{Encoding time.} The proposed system achieves a frame-wise
rate-distortion optimization during encoding. For a $1920 \times 1080$ video,
each frame requires 9 minutes on average. While far from being real-time, this
could be sped up through better implementation (\textit{e.g.} CUDA-based as in
\cite{mueller2022instant}) and by re-using a previous frame encoding as a better
initialization of the networks for the current frame.

\paragraph{High rate performance.} As detailed previously, although
Cool-chic video performance is compelling at low rates, it deteriorates at
higher rates. The degradation in performance can be attributed to inadequate
prediction. Surprisingly, at high rates, the prediction tends to be poor
meaning that the reference frames are not fully exploited. Indeed, at high
rates, the reference frames are of higher quality, containing more texture and
less blurring compared to low rates. Consequently, calculating optical flows on
such frames becomes more challenging. During training, the system may fall into
local minima by relying more on the residual rather than the optical flows. One
potential approach to enhance the system would be to prioritize the generation
of accurate predictions at high bitrates during the encoding process.


\section*{Conclusion}
This paper presents an overfitted video codec. It is designed by supplementing
Cool-chic with an inter coding module performing motion compensation and
residual coding. The resulting video codec shows promising compression
performance, close to AVC for both low-delay and random access coding.
Although there is still room for improvement, the proposed
decoder stands out for its low complexity with around 800 parameters and 900
multiplications per decoded pixel. We believe that this low decoding complexity
could ease the adoption of learned codecs for practical use.


\section{References}
\bibliographystyle{IEEEbib}
\bibliography{refs}

\end{document}